\documentstyle[twocolumn,prb,aps,epsfig]{revtex}
\begin{document}
\draft
\draft\twocolumn[\hsize\textwidth\columnwidth\hsize\csname
@twocolumnfalse\endcsname
\title{A Raman Study of Morphotropic Phase Boundary in
PbZr$_{1-x}$Ti$_x$O$_3$ at low temperatures}
\author{ K. C. V. Lima, A. G. Souza \thanks{
Corresponding Author: A. G. Souza Filho, FAX 55 (85) 2889903,
E-mail: agsf@fisica.ufc.br}, A. P. Ayala, J. Mendes Filho,\\ P. T.
C. Freire, and F. E. A. Melo}
\address{Departamento de F\'{\i}sica, Universidade Federal do
Cear\'{a}, Caixa Postal 6030,\\ Campus do Pici, 60455-760
Fortaleza, Cear\'{a}, Brazil}
\author{E. B. Ara\'ujo and J. A. Eiras}
\address{Departamento de F\'{\i}sica, Universidade Federal de S\~ao Carlos,
Caixa Postal 676,\\ 13565-670 S\~ao Carlos SP - Brazil}
\date{\today}

\maketitle

\begin{abstract}
Raman spectra of PbZr$_{1-x}$Ti$_x$O$_3$  ceramics with titanium
concentration varying between 0.40 and 0.60 were measured at 7 K.
By observing the concentration-frequency dependence of vibrational
modes, we identified the boundaries among rhombohedral,
monoclinic, and tetragonal ferroelectric phases. The analysis of
the spectra was made in the view of theory group analysis making
possible the assignment of some modes for the monoclinic phase.

\noindent Pacs number: {77.84.Dy, 77.84.-s, 77.80.Bh}
\end{abstract}

] \narrowtext

\bigskip

\newpage
\section{Introduction}

The PbZr$_{1-x}$Ti$_{x}$O$_3$ (PZT) solid-solution system contains
several compositions that are suited to important technological
applications in the electronic field such as piezoelectric
transducers, pyroelectric detectors, and non-volatile
ferroelectric memories. The main applications of PZT have been as
transducers that are fabricated using compositions closely related
to the Morphotropic Phase Boundary (MPB)\cite{bj}. For this
reason, MPB has been widely investigated from both experimental
techniques\cite{some} and theoretical approaches\cite{haun}. In
this region, PZT exhibits outstanding electromechanical properties
which have been attributed to the compositional fluctuations being
they treated within the framework of a phase coexistence
theory\cite{isupov}. Following this, a detailed nature of physical
properties for compositions close to MPB was not well established.

The recent discovery of a new monoclinic ferroelectric phase by
Noheda et al.\cite{bn} has shed a new light on the understanding
of the dielectric and piezolelectric enhancement for compositions
in the vicinity of MPB. In fact, the monoclinic distortion was
interpreted as either a condensation along one of the (110)
directions of the local displacements present in the tetragonal
phase, \cite{rguo} or as a condensation along one of the (100)
directions of the local displacements present in the rhombohedral
phase.\cite{corker} This monoclinic phase exhibited by PZT at low
temperatures would be the first example of a ferroelectric
material with P$_x^2$=P$_y^2$$\neq $P$_z^2$, where P$_x^2$,
P$_y^2$, P$_z^2$ $\neq$ 0. \cite{bn} Hence, the monoclinic
structure can be considered as a derivative form from both
tetragonal and rhombohedral phases by representing a link between
them. This model provides a microscopic picture where such a
striking electromechanical response close to the MPB region
\cite{rguo} is associated with the monoclinic distortion. A
successful explanation for the large piezoelectricity found in PZT
ceramics near the MPB was recently obtained by means of
first-principle calculations\cite{bell}. The piezoelectric
coefficients were calculated by considering the rotation of the
polarization vector in the monoclinic plane that is the unique
characteristic of monoclinic phase when compared with all other
ferroelectrics\cite{bell}. Also, the stability of the monoclinic
phase was recently described by means of phenomenological
thermodynamic studies within the framework of Landau-Devonshire
theory\cite{agsf} by considering the linear coupling between the
polarization and the monoclinic distortion $\beta$-90$^o$.

In spite of all investigations of the monoclinic phase in PZT
system, Raman studies performed on this new phase
\cite{agsf,franti} were limited to only two Ti concentration.
However, information on the phase transitions for other PZT
compositions close to the MPB is strongly desirable and appears as
an interesting investigation. In this way, the purpose of this
work is to investigate the extension of monoclinic phase for
compositions close to the MPB through Raman spectroscopy at low
temperatures. A careful analysis of the data allowed us to
determine the extension of monoclinic phase based on the phonon
behavior at low temperatures.

\section{Experimental}

Samples of PbZr$_{1-x}$Ti$_{x}$O$_3$ (with 0.40 $\leq x \leq$
0.52) were obtained through the solid-state reaction from 99.9
$\%$ pure reagent grade PbO, ZrO$_2$ and TiO$_2$ oxides. The
starting powders and distilled water were mixed and milled during
3.5 hours for powder homogenization. The mixture was calcined at
850 $ ^oC$ for 2.5 h. It was pressed at 400\thinspace MPa giving
rise to PZT ceramics disks with 10\thinspace mm of diameter and
5\thinspace mm of thickness. Finally, the disks were sintered at
temperature of 1250 $^oC$ during 4h and an excellent ceramics
homogeneity was obtained. The sintering atmosphere, which consists
of a covered alumina crucible, was enriched in PbO vapor using
PbZrO$_3$ powder around the disks in order to avoid significant
volatilization of PbO.

MicroRaman measurements were performed using a T64000 Jobin Yvon
Spectrometer equipped with an Olympus microsocope and a N$_2$-
cooled CCD to detect scattered light. The spectra were excited
with an Argon ion Laser ($ \lambda$ = 514.5 nm). The spectrometer
slits were set to give a spectral resolution of better than 2
cm$^{-1}$. A Nikon 20x objective with focal distance 20 mm and
numeric apperture N.A. = 0.35 was employed to focus the laser beam
on the polished sample surface. Low temperature measurements were
performed using an Air Products closed-cycle refrigerator which
provides temperatures ranging from 7 to 300 K. A Lakeshore
controller was used to control the temperature with precision of
order of $\pm$0.1 K.

We have made a systematic study of the laser-induced heating on
the samples in order to estimate the temperature trough the
Stokes/anti-Stokes ratio and improve a better signal/noise ratio.
The samples close to the MPB present a strong elastic scattering
which leads to a calculation of temperature by means of the
Stokes/anti-Stokes ratio not reliable. Then, we used a
PbZr$_{0.98}$Ti$_{0.02}$O$_3$ sample whose spectra can be easily
fitted and the temperature estimate was reproductive. Following
our observations we used a laser power of order of 0.1 mW.

\section{Result and Discussion}

In Fig. 1, Raman spectra for several PZT compositions varying from
0.40 to 0.60 are displayed. From x = 0.40 to x = 0.46, the spectra
remain exactly the same or change slightly, not only in frequency
but also in relative intensity for all modes. At 7~K, it is
expected a phase transition from rhombohedral low temperature
(R$_{LT}$) to rhombohedral high temperature (R$_{HT}$) phase. The
boundary between R$_{LT}$ and R$_{HT}$ has been established by
Jaffe et al\cite{bj} for temperatures higher than 300~K. An
extension of this boundary to low temperatures was made by Amim et
al\cite{amin} for composition x=0.40, and more recently by Noheda
et al\cite{lanl} for x=0.42.

The R$_{LT}$ phase, whose space group is C$_{3v}^6$, is
characterized by opposite rotations of adjacent oxygen octahedra
along the [111] polar axis. Due to this fact, the unit cell of
R$_{LT}$ phase has a volume twice the unit cell of R$_{HT}$ whose
space group is C$_{3v}^5$. From the diffraction point of view,
this transition is detected through the appearance of very weak
superlattice peaks which are originated just from the doubling of
the unit cell\cite{amin}. Concerned with the vibrational
properties, the dispersion relation for R$_{LT}$ phase is obtained
by folding the dispersion curves of R$_{HT}$ phase. Thus, the
observation of additional Raman modes is expected since wave
vectors belonging to the zone boundary have after the zone
folding, pseudomenta equivalent to q=0. For Zr-rich PZT, this
phase transition was observed from Raman spectroscopy by analyzing
the low frequency modes located at about 62 and 68~cm$^{-1}$ which
are, according to El-Harrad et al\cite{becker}, exclusive features
of the R$_{LT}$ and R$_{HT}$ phase, respectively. The mode at
62~cm$^{-1}$ is initially a zone boundary mode that becomes a zone
center active mode due to the doubling of the unit cell. In the
vicinity of the MPB, this kind of analysis is somewhat complicated
due to the overlapping of the bands which difficult the fitting
procedure and the identification of phase transition. In the
intermediate frequency region (120 $\leq$ $\omega$ $\leq$ 400
cm$^{-1}$, shown in Fig. 1) significant changes are not expected
in the vibrational modes since they are closely related to the
intrinsic modes of octahedral units such as stretching, torsion,
and bending.

Upon increasing Ti concentration, the spectra for 0.47 and 0.50
exhibits new features. Two of them are remarkable. First, the
intensity of the mode located at about 240~cm$^{-1}$ (marked with
a solid arrow) in rhombohedral phase decreases entering into
background for concentrations higher than x~=~0.48. Second, the
mode at about 280~cm$^{-1}$ (marked with a dot arrow) for x=0.40
presents a splitting resulting in a doublet mode. These spectral
changes were interpreted as due to the rhombohedral~-~monoclinic
phase transition. Thus, the boundary between R$_{LT}$ and
monoclinic structure is closely related to composition 0.46.
Noheda et al\cite{lanl} observed that PZT with x=0.46 is in a
monclinic structure down to 20~K. Our results show that this
composition is justly in the rhombohedral - monoclinic transition
region. Since the boundary between R$_{HT}$ and monoclinic
structure is almost vertical, we believe that a small deviation in
composition accounts for the observed disagreement between our
results and those reported in Ref. \cite{lanl}.

In the first analysis, there are no clear changes in the Raman
spectra from x=0.48 to x=0.60. However, the monoclinic-tetragonal
phase transition is expected around x=0.52\cite{lanl}. In order to
study the effects of the phase transition in the phonon spectra we
constructed the frequency versus x ($\omega$ $vs.$ x) plot where
the transitions can be observed in details. The frequencies were
obtained by deconvoluting the spectra using a set of curves with
lorentzian shapes. The number of peaks used to fit spectra were
determined by means of group theory analysis that will be
discussed in a forthcoming paragraph. In Fig. 2, we shown the
deconvoluted spectrum for tetragonal PbZr$_{0.60}$Ti$_{0.40}$O$_3$
following the procedure and assignment earlier
reported\cite{franti} where observed frequencies are in good
agreement with those found there.

In order to understand the tetragonal~-~monoclinic transition, let
us first discuss the mode symmetry related to tetragonal phase in
PbTiO$_3$. When the cubic phase transforms into tetragonal, the
T$_{2u}$ silent mode transforms into B$_1$ $\oplus$ E irreducible
representation of the C$_{4v}$ and a degeneracy breaking of this
mode is expected\cite{burns}. Besides its Raman activity in
C$_{4v}$ symmetry, this mode has been called a {\it silent} mode
and its splitting was not observed for PbTiO$_3$ at room
temperature\cite{burns}. However, when Ti is replaced by Zr, the
splitting is observed at low temperatures. The member of the
doublet with higher frequency is assigned as the B$_1$
mode\cite{franti} and the value $\omega _B{_1}$ - $\omega_E$
increases in the vicinity of MPB. This observation was made by
Frantti et al\cite{franti} who studied this feature from
compositions x=0.49, 0.50, 0.60, 0.70, 0.80 and 0.90. Here, we
extend this study inserting intermediate compositions making
possible a detailed description of B$_1$ mode when
monoclinic~-~tetragonal transition take place. To describe the
phase transition in details, we have constructed, based on the
group theory analysis, the $\omega$ $vs.$ x plot shown in Fig. 3a.
The point group of monoclinic phase is C$_s$ where all the
irreducible representation (A$^{\prime}$ and A$^{\prime \prime}$)
are Raman active. The C$_s$ group is subgroup of both C$_{3v}$ and
C$_{4v}$ and the correlation between them and O$_h$ can be
summarized in the scheme showed in Fig. 4. Following that scheme,
we observe that B$_1$ and E modes belonging to the tetragonal
phase transforms into A" and A$^{\prime}$ $\oplus$ A$^{\prime
\prime}$, respectively. Therefore, we have used three modes in the
monoclinic phase to fit the region around 280~cm$^{-1}$. It is
interesting to note that highest frequency member of doublet
A$^{\prime}$ $\oplus$ A$^{\prime \prime}$ presents the same
behavior early reported by Frantti et al\cite{franti}.

We also plotted in Fig. 3b our results and those earlier
reported\cite{franti,burns} for several compositions in a small
frequency region. In this figure we can observe that B${_1}$ mode
decreases in frequency from x=1.0 to x=0.60 and increases from
x=0.60 to x=0.52. Below x=0.52, it decreases(increases) for Ti
content from  0.52 to 0.50 (0.48 to 0.47). The
tetragonal~-~monoclinic phase transition can be observed by
analyzing the splitting $\Delta \omega$ ($\omega_{B_1}$-$\omega_E$
in tetragonal phase and $\omega_{A^{\prime \prime}}$-the highest
frequency member of A$^{\prime}$ $\oplus$ A$^{\prime \prime}$  in
monoclinic phase). This splitting in the tetragonal phase ($x$
$\ge$ 0.52) decreases when Ti content increases while it presents
the same behavior exhibited by B$_1$ mode in the monoclinic region
(Fig. 3c). Both changes in B$_1$ and in $\Delta \omega$ indicate
the transition between monoclinic and tetragonal symmetry. Noheda
et al\cite{lanl} reported that composition x=0.52 presents a
tetragonal symmetry at 20~K. Our results at 7~K indicate that this
composition is in a monoclinic phase. In principle, our results do
not differ from those of Ref. \cite{lanl} and we believe that
either a small deviation in $x$ or the extension of
monoclinic~-~tetragonal boundary can account for the minor
disagreement between their and our results. For the sake of
completeness, the frequencies of doublet mode depicted in Fig. 3b
for several compositions are listed in Table I.

A similar analysis based on the group theory can be performed to
describe the R$_{HT}$~-~monoclinic phase transition observed
around x=0.46. When the cubic (O$_h$) phase transforms into
rhombohedral C$_{3v}$ phase, the T$_{2u}$ silent mode transforms
into A$_2$ $\oplus$ E irreducible representation being A$_2$
without Raman activity. For this reason, there is a single peak at
about 280~cm$^{-1}$ for compositions with rhombohedral phase. By
observing the correlation between the C$_{3v}$ and C$_s$ depicted
in Fig. 4 we can observe the splitting of this mode into three new
modes, i.e., A$^{\prime}$ $\oplus$ 2A$^{\prime \prime}$ which are
in a perfect accordance when we describe the transition from
tetragonal side.

Let us now try to discuss the monoclinic phase based on the
assumptions of distortion of the unit cell by strain. The polar
axis in the monoclinic phase has a particular feature if it is
compared with that of tetragonal and rhombohedral phase. In fact,
it can not be determined only by symmetry and can be along any
direction within the monoclinic plane. This consideration was
introduced by first-principle calculation made by Bellaiche et
al\cite{bell} where the striking piezoelectric properties for
monoclinic compositions were successfully obtained. To the best of
our knowledge, there is no systematic theoretical study concerned
with what happens with optical phonons when PZT transition occurs.
In fact, there are only two studies which report the calculation
of vibrational frequency for PbTiO$_3$. Freire and
Katiyar\cite{freire} carried out such calculation just by
adjusting the parameters of a rigid ion model while Garcia and
Vanderbilt\cite{garcia96} performed it using the first-principle
calculation. Following the results of these latter authors the
tetragonal phase could change either to an orthorhombic or to a
monoclinic phase by means of the linear coupling between strain
and atomic displacement.

We are reminded of this result because the tetragonal phase in PZT
is very similar to that of in PbTiO$_3$ and the
monoclinic~-~tetragonal transition is marked by changes in the
B$_1$ symmetry mode. This is very interesting because this mode
involves only oxygen motion [O$_{1z}$-O$_{2z}$] where both oxygen
atoms move in opposite directions leading to the tetragonal
symmetry-breaking. However, this mode appears to be dependent of
the Zr/Ti ratio as demonstrated by Frantti et al\cite{franti}
whose part of these results are reproduced in Fig. 3b. When the
temperature is kept constant, the variable responsible for the
transition is the Zr/Ti ratio. It is well-known that the
tetragonal strain, c$_t$/a$_t$, decreases when Zr content
increases. Taking into account this fact, the unit cell can only
be distorted by an in-plane strain. Considering that the strain
can transform according to the irreducible representation of
C$_{4v}^1$ space group, where the E distortion leads to a
monoclinic symmetry and B$_1$ and B$_2$ to an orthorhombic one
with axes parallel (B$_1$) or rotated about 45$^o$ (B$_2$) with
respect to the tetragonal base. In this way, a B$_2$ distortion
should be the transition mechanism since the actual monoclinic
phase is just characterized by a rotation of 45$^o$ in tetragonal
plane. Also, the monoclinic phase can be seen as a
pseudo-orthorhombic one due to the very small monoclinic angle
$90^o-\beta$ whose origin can be in the delicate balance between
temperature- and concentration-induced strain. The lattice
dynamical study made by Garcia et al\cite{garcia96} is limited to
the edge of the PZT phase diagram where the doubling of the unit
cell was not considered. Finally, the observed behavior of optical
phonons as a function of concentration seems to be confirmed by
means of first-principle calculation considering the effects of
Zr/Ti ratio. This can provide a better understanding of the phonon
related phase transition in PZT as well as other perovskite
systems with similar MPB.

\section{conclusions}

In summary, the behavior of the optical modes for
PbZr$_{1-x}$Ti$_{x}$O$_3$ with $x$ varying from x=0.40 to x=0.60
were reported and discussed. The observed changes were attributed
to the phase transitions. At 7~K, we were able to observe the
transitions rhombohedral~-~monoclinic~-~tetragonal. Our results
are in accordance with those recently reported by Noheda et
al.\cite{lanl} and Frantti et al.\cite{franti} which studied the
stability of monoclinic phase using high resolution synchrotron
X-ray powder diffraction technique. Both transitions we have
reported are studied in the view of group theory analysis where
some optical phonons in the monoclinic phase were assigned. Very
of interest is the behavior of B$_1$ mode that has an unusual
concentration dependence in the vicinity of tetragonal-monoclinic
phase transition. Further studies in PZT single crystal  with
monoclinic phase will be need in order to assign each member of
doublet A$^{\prime}$ $\oplus$ A$^{\prime \prime}$.

\vskip0.5truecm\noindent {\bf Acknowledgements}~~--~~ The authors
wish to acknowledge Dr. I. Guedes for valuable discussions related
to this work. One of us, A. G. Souza Filho wishes to acknowledge
the fellowship received from FUNCAP.  Financial support from
FUNCAP, CNPq, FAPESP and FINEP is gratefully acknowledged.

\begin{figure}[htbp]
\caption{Raman spectra of PbZr$_{1-x}$Ti$_{x}$O$_3$ ceramics
recorded at 7~K. The numbers stand for Ti concentration.}
\end{figure}

\begin{figure}[htbp]
\caption{Raman spectrum of PbZr$_{0.40}$Ti$_{0.60}$O$_3$
illustrating the fitting procedure employed to deconvolute it in a
set of Lorentzian curves.}
\end{figure}

\begin{figure}[htbp]
\caption{a) Variation of the frequency of some Raman modes as a
function of concentration recorded at 7~K; b) a small frequency
region shown in a) is depicted and; c) the splitting $\Delta
\omega$ ($\omega_{B_1}$-$\omega_E$ in tetragonal phase and
$\omega_{A^{\prime \prime}}$-the highest frequency member of
A$^{\prime}$ $\oplus$ A$^{\prime \prime}$  in monoclinic phase.
The dotted lines in a) and b) represent the transition region
among rhombohedral (R), monoclinic (M), and tetragonal (T). The
vertical dotted line in c) is only a guide for eyes. The labels
stand for symmetry modes in different phases. The open-circles and
open-square are data from Ref. \cite{franti} and Ref. \cite{burns}
where all solid-circles are data from the present work.}
\end{figure}

\begin{figure}[htbp]
\caption{Correlation table for some modes belonging to the
monoclinic (C$_s$), tetragonal (C$_{4v}$), and rhombohedral
(C$_{3v}$) phase originated from T$_{2u}$ in cubic (O$_h$) phase.}
\end{figure}

\begin{center}
\begin{table}
\caption{The values of each member of the doublet showed in Fig.
3b for various compositions of PZT recorded at 7~K. The
frequencies are in units of cm$^{-1}$. The label $a$ and $b$
stands for results obtained in this work and in Ref.
\cite{franti}, respectively.}
\begin{tabular}{ccc}
PbZr$_{1-x}$Ti$_{x}$O$_3$& This work & Frantti et al\cite{franti}
\\ \hline

PbZr$_{0.525}$Ti$_{0.475}$O$_3$  & 270,287 &  \\

PbZr$_{0.52}$Ti$_{0.48}$O$_3$  &272,294 & \\

PbZr$_{0.51}$Ti$_{0.49}$O$_3$  &&267,293   \\

PbZr$_{0.50}$Ti$_{0.50}$O$_3$ &268,293&267,293 \\

PbZr$_{0.49}$Ti$_{0.51}$O$_3$ &267,292 & \\

PbZr$_{0.48}$Ti$_{0.52}$O$_3$  & 268,291 &  \\

PbZr$_{0.45}$Ti$_{0.55}$O$_3$  &  267,289 &  \\

PbZr$_{0.42}$Ti$_{0.58}$O$_3$  &  271,287  & \\

PbZr$_{0.40}$Ti$_{0.60}$O$_3$  & 272,288 &271,286 \\

PbZr$_{0.30}$Ti$_{0.70}$O$_3$  & &277,287 \\

PbZr$_{0.20}$Ti$_{0.80}$O$_3$  & &282,289 \\

PbZr$_{0.10}$Ti$_{0.90}$O$_3$  &  &287,291\\
\end{tabular}
\end{table}
\end{center}
\end{document}